%% file: main.tex
\documentclass[journal]{IEEEtran}

\ifCLASSINFOpdf
\else
   \usepackage[dvips]{graphicx}
\fi
\usepackage{url}
\usepackage{amsmath, amssymb, cite, bbold} 
\usepackage{amsfonts}

\usepackage{enumitem}
\usepackage{calc,amsfonts,amssymb,amsmath,bm,url,color,theorem,graphicx,cite,epstopdf,nicefrac,bbold}

\usepackage[algoruled,linesnumbered]{algorithm2e}
\usepackage{graphicx}
\usepackage{subcaption}
\usepackage{bm}
\usepackage{booktabs}

\input{def}

\definecolor{orange}{RGB}{255,107,0}
\definecolor{green}{RGB}{50,170,50}
\definecolor{magenta}{RGB}{255,0,255}

\newtheorem{Lemma}{Lemma}
\newtheorem{Prop}{Proposition}

\theorembodyfont{\rmfamily}

\begin{document}

\title{Domain-Factored Untrained Deep Prior for Spectrum Cartography}

\author{Subash Timilsina, Sagar Shrestha, Lei Cheng, and Xiao Fu
\thanks{This work of S. Timilsina, S. Shrestha, and X. Fu was supported in part by the National Science Foundation (NSF) under Projects NSF ECCS-2024058 and NSF CCF-2210004. {\it Corresponding author: Xiao Fu}. }
\thanks{The work of L. Cheng was supported in part by the National Natural Science Foundation of China under Grant 62371418.}
\thanks{S. Timilsina, S. Shrestha, and X. Fu are with Oregon State University, Corvallis, OR 97331 USA.}
\thanks{L. Cheng is with Zhejiang University, Hangzhou, China.}}

\maketitle

\begin{abstract}

\textit{Spectrum cartography} (SC) aims to estimate the radio power map of multiple emitters over space and frequency using limited sensor data.  
Recent advances leverage learned \textit{deep generative models} (DGMs) as structural priors, achieving state-of-the-art performance by capturing complex spatial-spectral patterns.  
However, DGMs require large training datasets and may suffer under distribution shifts.
To address these limitations, we propose a \textit{training-free} SC approach based on \textit{untrained neural networks} (UNNs), which encode structural priors through architectural design.  
Our custom UNN exploits a spatio-spectral factorization model rooted in the physical structure of radio maps, enabling low sample complexity.  
Experiments show that our method matches the performance of DGM-based SC without any training data.
\end{abstract}


\IEEEpeerreviewmaketitle

\section{Introduction}
This letter revisits the \textit{spectrum cartography} (SC) task in \cite{boccolini2012wireless, romero2017learning, zhang2020spectrum, shrestha2022deep, mateos2009spline, kim2013cognitive, sun2022propagation, sun2024integrated, sun2024tensor}.
The goal is to construct a radio frequency (RF) interference propagation map over space and frequency from limited measurements sent by sparsely deployed sensors.
Earlier works proposed to use certain handcrafted structural constraints of the radio map (e.g., low-rank, sparsity, and smoothness) to regularize this clearly ill-posed inverse problem; see, e.g., \cite{romero2017learning,bazerque2011group,zhang2020spectrum}. To handle more complex RF environments, particularly where heavy shadowing happens, 
deep generative models (DGMs) were used to learn the structural characteristics of radio maps; see, e.g.,
\cite{teganya2020data, shrestha2022deep, timilsina2023quantized, han2020power}. 
Specifically, a deep neural network (DNN)-based representation is learned using offline training data. Then, the DNN is used as a structural constraint in the inverse problem solving process of SC.
Due to DNN's expressiveness, such ``learned priors'' can capture detailed and complex characteristics of radio maps, offering state-of-the-art performance of SC \cite{shrestha2022deep, timilsina2023quantized}.

However, learning a DGM consumes a large number of training data, which is not always available.
In addition, mismatches between the training data and actual radio map environments could be detrimental to the SC performance, which is particularly problematic when the environment is dynamic or fast changing. 

\noindent
\textbf{Contributions.}  
We propose using the \textit{untrained neural network} (UNN) approach \cite{ulyanov2018deep} for SC.  
While UNNs have shown promise in vision tasks by leveraging architectural bias without training, directly applying them to radio maps leads to high model and sample complexities.
To address this, we design a custom UNN based on a spatio-spectral factorization model, motivated by the approximate uncorrelatedness between spatial and spectral components over narrow bands \cite{fu2015factor, romero2017learning, zhang2020spectrum, fu2016power, bazerque2011group}.  
This factorization enables efficient representation of emitter-specific spatial components.  
We further exploit shared statistical patterns among emitters to reduce complexity.
Simulations demonstrate that our method performs comparably to learned DGM-based SC approaches---despite requiring no training data.

\begin{figure}[!t]
    \centering
    \includegraphics[width=0.99\linewidth]{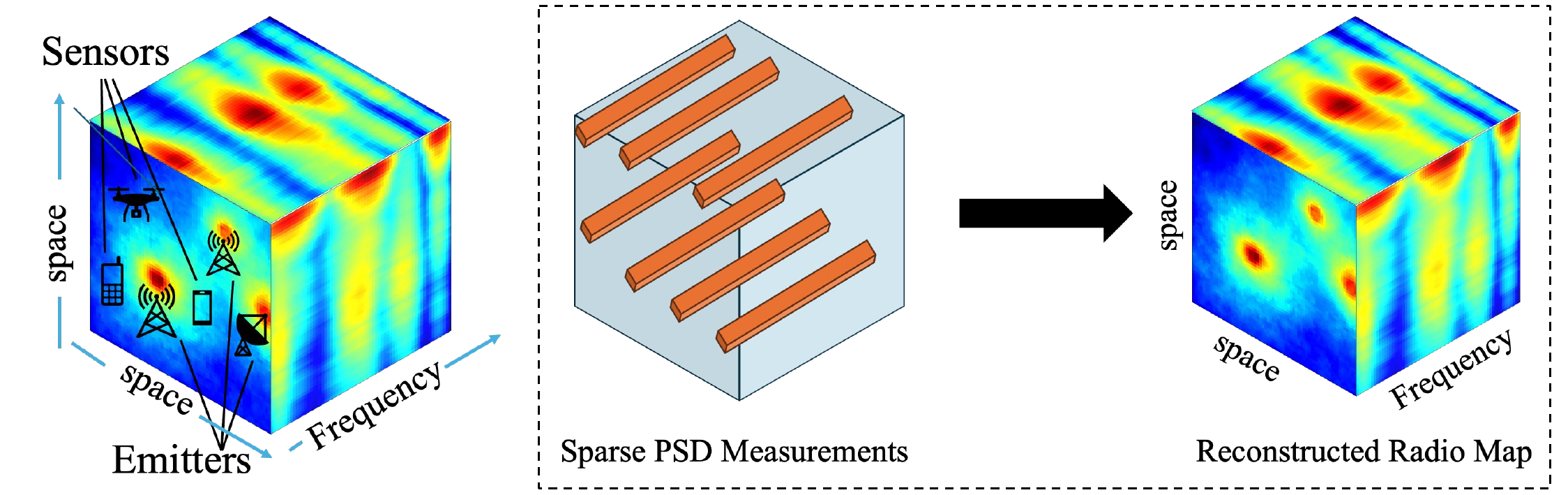}
    \caption{(Left) A spatio-spectral radio map. (Right) Problem setup and objective of SC; adapted from \cite{shrestha2022deep,timilsina2023quantized}.}
    \label{fig:radiomap_sc_illustration}
    \vspace{-0.25cm}
\end{figure}

\section{Problem Statement and Background}

We consider an SC scenario where $R$ emitters share $K$ frequency bins; see, e.g., \cite{bazerque2011group, teganya2020data, schaufele2019tensor, zhang2020spectrum, timilsina2023quantized, romero2017learning, han2020power}. 
The spatial region is divided into $I \times J$ grid cells, and $N$ sensors are deployed in this region. Power propagation of the $R$ emitters in the region over $K$ frequencies forms an $I \times J \times K$ radio map tensor $\tX \in \bbR^{I \times J \times K}$, where each entry of tensor $\tX(i, j, k)$ is the the \textit{power spectral density} (PSD) of signal received at location $(i, j)$ and frequency $k$; see Fig.~\ref{fig:radiomap_sc_illustration} (left). 
Let us denote $\bm \varOmega = \{ (i,j)~|~ i \in [I],~ j \in [J]\} \subseteq [I] \times [J]$ as the set of sensor locations, where $|\bm \varOmega| = N \ll IJ$. The sensor at $(i, j)$ collects $\tX(i, j, :)$, i.e., the PSD at location $(i, j)$, and transmits it to a fusion center.
The goal of SC is to recover $\tX$ from $\{\tX(i, j, :) \}_{(i, j) \in \bm \varOmega}$; see Fig. \ref{fig:radiomap_sc_illustration} (right). 

SC was tackled using many classical inverse problem solving techniques, e.g., spatial smoothness-based kernel interpolation \cite{boccolini2012wireless,bazerque2011group}, representation sparsity-based dictionary learning \cite{bazerque2010distributed} and compressive sensing \cite{jayawickrama2013improved, jayawickrama2014iteratively},
and low-rank based matrix/tensor completion \cite{zhang2020spectrum, sun2024integrated}.
However, the effectiveness of these methods decreases in complex RF environments, e.g., those with heavy shadowing. This is because assumptions such as spatial/spectral smoothness or low-rank structures of radio maps could be grossly violated in such cases \cite{shrestha2022deep,timilsina2023quantized}.

More recently, deep learning-based SC methods have been proposed to tackle these challenges (e.g., \cite{teganya2020data, shrestha2022deep, timilsina2023quantized, han2020power}). These approaches learn a deep generative model (DGM) to represent radio maps as $\tX = {\cal G}_{\bm \theta}(\bm z)$, where $\bm z \in \mathbb{R}^d$ is a latent code and ${\cal G}_{\bm \theta}$ is trained on data reflecting the target environment. DGMs can capture complex radio map structures, even under severe shadowing. However, they require large amounts of training data \cite{shrestha2022deep}, and performance suffers when there is a domain gap---e.g., due to environmental changes. In such cases, retraining may be needed, limiting their practicality for {\it in situ} sensing.

\section{Proposed Approach}
\label{sec:proposed approach}

\subsection{Preliminaries of UNNs}
A way to circumvent training DGMs yet retaining representation expressiveness is to use UNNs. 
UNNs were first introduced in the computer vision literature \cite{ulyanov2018deep}. It was observed that the inherent structure of a convolutional neural network, when carefully designed, can act as a strong prior for natural images; i.e., the expression $\tX={\cal G}_{\bm \theta}(\bm z)$ holds for unknown $\bm \theta$ and a certain $\bm z$ under careful construction of ${\cal G}_{\bm \theta}$---and $\bm \theta$ can be learned to adapt any given data $\tX$. Using UNNs, the data recovery problems can be formulated as
\begin{align}\label{eq:unnlearn}
    \minimize_{ \bm \theta }~\left\| \tM \circledast \left(\tX - {\cal G}_{\bm \theta}(\bm z) \right)     \right\|_{\rm F}^2,
\end{align}
where $\bm z$ could be fixed, e.g., a Gaussian/uniform variable \cite{ulyanov2018deep,miao2021hyperspectral} or a learnable term \cite{li2024zero}.
The UNN-based approach attracted a lot of attention in vision, as it bypasses the need for training data while keeping impressive representation accuracy for complex image data across disciplines \cite{ulyanov2018deep, heckel2019deep, heckel2020denoising, gadelha2019shape, uezato2020guided,miao2021hyperspectral,li2024zero}.

\subsection{UNN Design for Radio Maps}

Directly applying existing vision-based UNNs (e.g., \cite{ulyanov2018deep, heckel2019deep, heckel2020denoising, gadelha2019shape, uezato2020guided, miao2021hyperspectral, li2024zero}) to represent the $I \times J \times K$ radio map $\tX$ via \eqref{eq:unnlearn} is challenging---$\tX$ is often too complex to model compactly. For instance, as noted in \cite{shrestha2022deep}, placing just $R = 5$ emitters in a $100 \times 100$ grid yields $8 \times 10^{17}$ combinations, not accounting for other parameters like power or frequency. Capturing such complexity may require a highly overparameterized UNN (i.e., $|\bm \theta| \gg IJK$, where $|\bm x|$ denotes the cardinality of $\bm x$) to satisfy $\tX = {\cal G}_{\bm \theta}(\bm z)$ for arbitrary $\tX$ drawn from the distribution $ {\cal D}$. As a result, solving \eqref{eq:unnlearn} can demand an impractically large number of sensor measurements.

We design a custom UNN structure based on the classical spatio-spectral factorization model of radio maps  \cite{bazerque2011group, romero2017learning, zhang2020spectrum, shrestha2022deep, timilsina2023quantized}. Under this model, for a region with $R$ emitters, we have
\begin{align}\label{eq:radio_model}
\tX = \sum_{r=1}^R \bm S_r \circ \bm c_r,
\end{align}
where $\circ$ denotes the outer product, and $\bm c_r \in \bbR^K$ and $\S_r \in \bbR^{I \times J}$ are the PSD and the {\it spatial loss field} (SLF) of emitter $r$, respectively. The SLF captures the spatial power propagation pattern of an emitter.
The model is considered valid if the bandwidth of interest is not too wide compared to the central frequency \cite{fu2016power,fu2015factor,bazerque2011group,romero2017learning}; see Fig. \ref{fig:unn_model}.

As argued in \cite{shrestha2022deep,timilsina2023quantized}, the modeling difficulty of radio maps mainly lies in its SLFs. Hence, instead of using a large UNN to represent $\tX$, our idea is to use a ``frugal'' UNN to represent $\bm S_r$ for all $r$--- converting a 3D tensor modeling problem to a 2D SLF modeling one. Note that each $\bm S_r$ corresponds to only one emitter, and thus it also has a much smaller ``state space'' relative to $\tX$ (only $10^4$ emitter positions---as opposed to $8\times 10^{17}$---exist with a single $\bm S_r$ in the previous example).
Therefore, if one designs UNNs to represent $\bm S_r$'s, it is expected to be more parameter-efficient than using existing UNNs like those in \cite{ulyanov2018deep, heckel2019deep, heckel2020denoising, gadelha2019shape, uezato2020guided, miao2021hyperspectral, li2024zero} to model the entire $\tX$.

\subsection{Architecture of UNN for SLFs}
To model $\S_r$ succinctly, the UNN should have as few parameters as possible, enabling SC with fewer samples. At the same time, it must be expressive enough to capture complex RF environments---striking this balance is nontrivial.

\begin{figure}[t!]
    \centering
    \includegraphics[width=.88\linewidth]{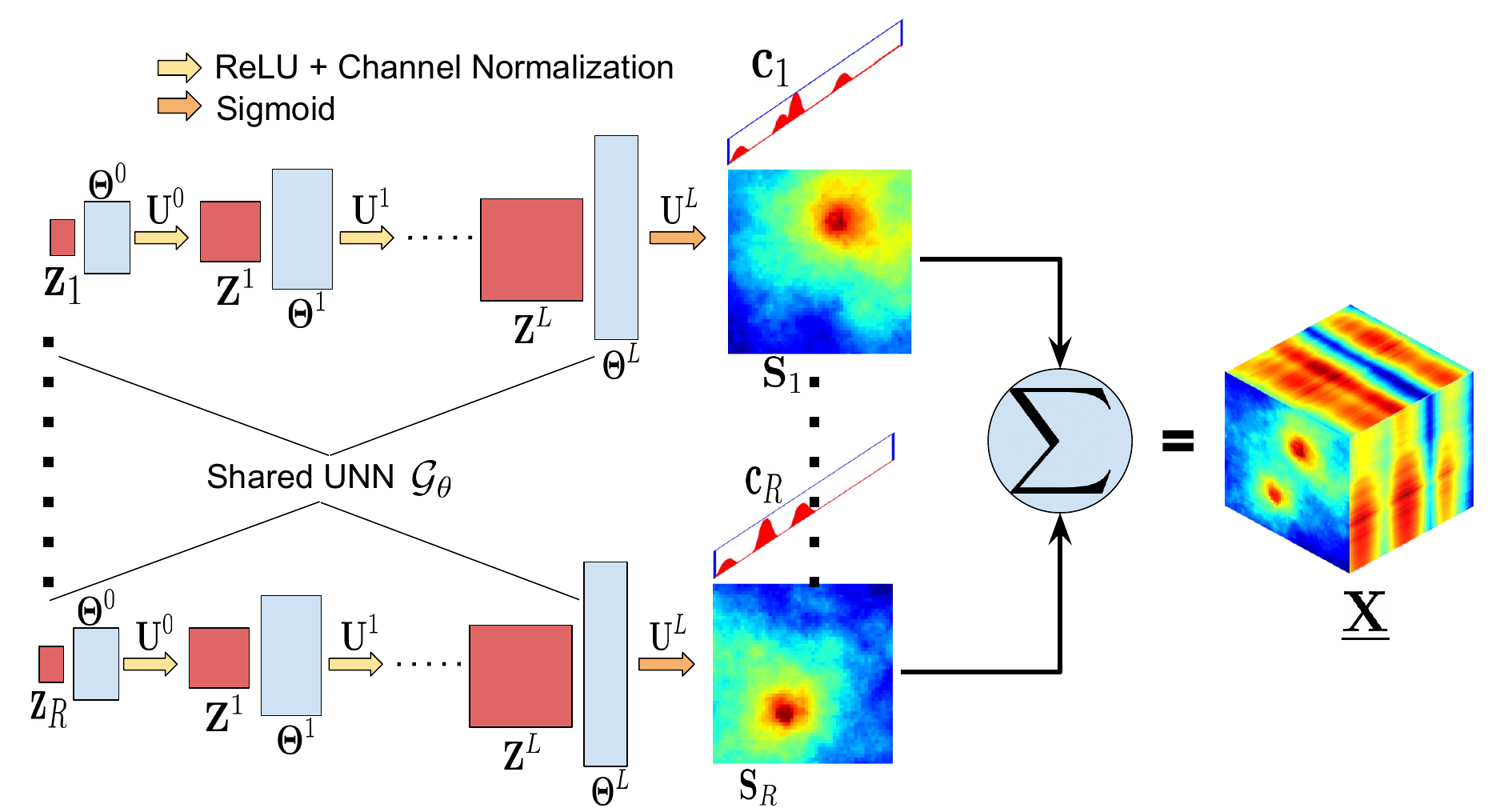}
    \caption{Proposed custom UNN design for radio maps.}
    \label{fig:unn_model}
\end{figure}

Notice that it is reasonable to assume that $\bm S_r$ for $r=1,\ldots,R$ follow the same distribution.
Therefore, $\bm S_r={\cal G}_{\bm \theta}(\bm z_r)$ for all $r$ can share the same generative model ${\cal G}_{\bm \theta}$ and only distinguish by their latent codes $\bm z_r$ (see \cite{shrestha2022deep,timilsina2023quantized}).
This already reduces model complexity substantially---another benefit of the factorization model in \eqref{eq:radio_model}.
To design ${\cal G}_{\bm \theta}$,
we use a decoder architecture inspired by \cite{heckel2019deep}.
The deep decoder structure \cite{heckel2019deep} often uses fewer parameters than other UNN architectures, e.g., the U-net based one in \cite{ulyanov2018deep}---which, together with our parameter-sharing design, further reduces the sample complexity for SC.
The architecture at layer $i+1$ (except for the final layer) can be written as
\begin{align} \label{eq:deep_dec_arch}
    \Z^{i+1} = {\rm Cn}( {\rm ReLU}(\U^{i} \Z^{i} \bm \Theta^{i}) ),
\end{align}
where $\Z^{i} \in \bbR^{D_i^2 \times k_i}$ represents the output of $i$th layer of the UNN, with $D_i^2$ and $k_i$ representing the spatial and the channel dimensions, respectively.
Similarly, $\bTheta^i \in \bbR^{k_i \times k_{i+1}}$ represents the parameters of UNN in the $i$th layer, and we have $\btheta = \{\bTheta^0 , \dots, \bTheta^L\}$. 
In implementation, $\Z^i, \forall i$ is a third-order tensor of shape $D_i \times D_i \times k_i$, and $\bTheta^i$ represents the parameters of convolution operation.
However, we use the tensors' matrix unfolding with a size of $D_i^2 \times k_i$ for the ease of analysis (see \cite{sidiropoulos2017tensor} for details of unfolding).
The input to the UNN is $\Z^0 \in \bbR^{D_0^2 \times 1}$ (i.e., the matrix unfolding version of $\z$), and the output is an SLF, i.e., $\S \in \bbR^{I \times J}$ generated as
$    \S = {\rm mat}({\rm Sigmoid}(\Z^{L} \bm \Theta^{L})).
$
This implies that $I = J = D_L$ and $k_0 = k_{L+1} = 1$. 
Fig.~\ref{fig:unn_model} shows the domain-factored and deep decoder-shared UNN designed for radio maps.
Note that directly modeling 3D tensors with a ``naive UNN'' with the deep decoder architecture without using the structure in \eqref{eq:radio_model} could lead to significantly more complex UNNs.

\subsection{Objective Function and Recoverability}\label{sec:recoverability}

In this section we denote the ground-truth radio map by $\gtX$.  
Each sensor at $(i,j)$ measures $\gtX(i,j,:)$ and transmits  
$\Y(i,j,:) = h(\gtX(i,j,:))\in\bbR^{K}$ to the fusion center, where  
$h(x)=\log(x+a)$ with $a>0$ compresses the dynamic range and mitigates numerical issues \cite{timilsina2023quantized}.  
With our UNN, radio map recovery is posed as
\begin{align}\label{eq:full_precision}
 \underset{\btheta, \Z, \C\geq \bm 0 }{\minimize} &~ 
 \Bigg\| \tM \circledast  \left(\tY - h \left( \sum_{r=1}^R {\cal G}_{\btheta}(\z_{r} ) \circ {\bm c}_r \right) \right) \Bigg\|_{\rm F}^2 
\end{align}
where $\Z=[\z_1,\ldots,\z_R]$ are latent codes and  
$\C=[\bm c_1,\ldots,\bm c_R]$ are PSDs (hence non-negative).  
For quantized data, replace \eqref{eq:full_precision} by the MLE of \cite{timilsina2023quantized} (see supplementary material).

The formulation in \eqref{eq:full_precision} provably recovers $\gtX$ up to bounded error. To see this, we define a solution set ${\cal X}^{\rm UNN} \subset \bbR^{I \times J \times K}$ that can be represented by our
domain-factored UNN in  Fig.~\ref{fig:unn_model}
as ${\cal X}^{\rm UNN} = \{ \tX~|~\tX=\sum_{r=1}^R {\cal G}_{\bm \theta}(\z_r)\circ \bm c_r,~  {\cal G}_{\bm \theta}(\z_r) \in \cH^{(\gamma)}, \bm C\in \cC \},$
where $\cH^{(\gamma)} =  \{ {\cal G}_{\btheta}(\z) ~|~\|{\cal G}_{\bm \theta}(\z_r) \|_{\rm F} \leq \gamma,~ \btheta \in \cM, \z \in \cZ \}$, $\cC = \{ \C ~|~ \C \geq \bm 0, \| \bm c_r \|_2 \leq \kappa, \forall r \in [R] \}$, and 
$\cM =  \{ \btheta = [\bm \Theta^0 \ldots \bm \Theta^L], \bm \Theta^{i} \in \bbR^{k_i \times k_{i+1}} ~|~ \| \bm \Theta^{i} \|_{\sigma} \leq s, \| \bm \Theta^i \|_{2, 1} \leq b  \}$ and $\cZ = \{ \z \in \bbR^{D_0^2} ~|~ \| \z \|_{2} \leq a  \}$.
We characterize ${\cal X}^{\rm UNN}$'s complexity by the following lemma:
\begin{Lemma}\label{lemma:x_unn}
    The covering number of the $\epsilon$-net of ${\cal X}^{\rm UNN}$, denoted by $ {\sf N}({\cal X}^{\rm UNN}, \epsilon)$, can be upper bounded by
    \begin{align} \label{eq:cover_x_unn}
     &\log ( {\sf N}({\cal X}^{\rm UNN}, \epsilon) ) \leq  \nicefrac{R^3 (\kappa + \gamma) a^2 b^2 P \log(2 W^2) s^{2L - 2} L^3}{\epsilon^2}  \nonumber\\
     &+ R D_0^2 \log \left(  \nicefrac{6 R P a (\kappa + \gamma)}{\epsilon} \right ) + R K \log \left( \nicefrac{3 R \kappa (\kappa + \gamma)}{\epsilon} \right). 
    \end{align}
    
\end{Lemma}
The proof is in the supplementary material. 
Using the above, we show:
\begin{Prop}\label{prop:fp}
        Assume that $\bm \varOmega$ is uniformly sampled from $[I] \times [J]$ without replacement. Suppose that $\tX^\star\in {\cal X}^{\rm UNN}$ and that $\tX^{\ast} = \sum_{r=1}^R {{\cal G}_{\btheta^\star}(\z_r^{\ast})} \circ {\bc}_r^{\ast}$, where $(\bm \theta^\star,\{\bm z_r^\star,\bm c_r^\star\}_{r=1}^R)$ is any optimal solution of \eqref{eq:full_precision}.
      Then, the following holds with probability of at least $1 - \delta$: 
            \begin{align*}
                \frac{\|\tX^{\star} - \tX^{\natural}\|_{\rm F}}{\sqrt{IJK} }  \leq {\sf O} \left ( \frac{R}{\sqrt{N}} + \frac{(\log ( {\sf N}({\cal X}^{\rm UNN}, \epsilon) ) )^{\frac{1}{4}}}{\sqrt{K} N^{1/4}} \right )  + \nu^{\rm UNN} ,
            \end{align*}
        where $\nu^{\rm UNN} =\min_{{\tX}\in \cX^{\rm UNN}}\|\tX - \gtX \|_{\infty} $.
\end{Prop}
The proof is obtained by inserting Lemma~\ref{lemma:x_unn} into the proof of \cite[Theorem~2]{shrestha2022deep}, thereby omitted.
Our UNN keeps $L$, $s$, and $b$ small, tightening \eqref{eq:cover_x_unn} and improving sample complexity.  
Bounds on $\bm c_r$, $\bm z_r$, and $\bm\Theta^{i}$ are enforced with regularization; see the supplementary material for the quantized case.

\section{Experiments}\label{sec:experiments}
\begin{figure}[!t]
    \centering
    \includegraphics[width=.88\linewidth]{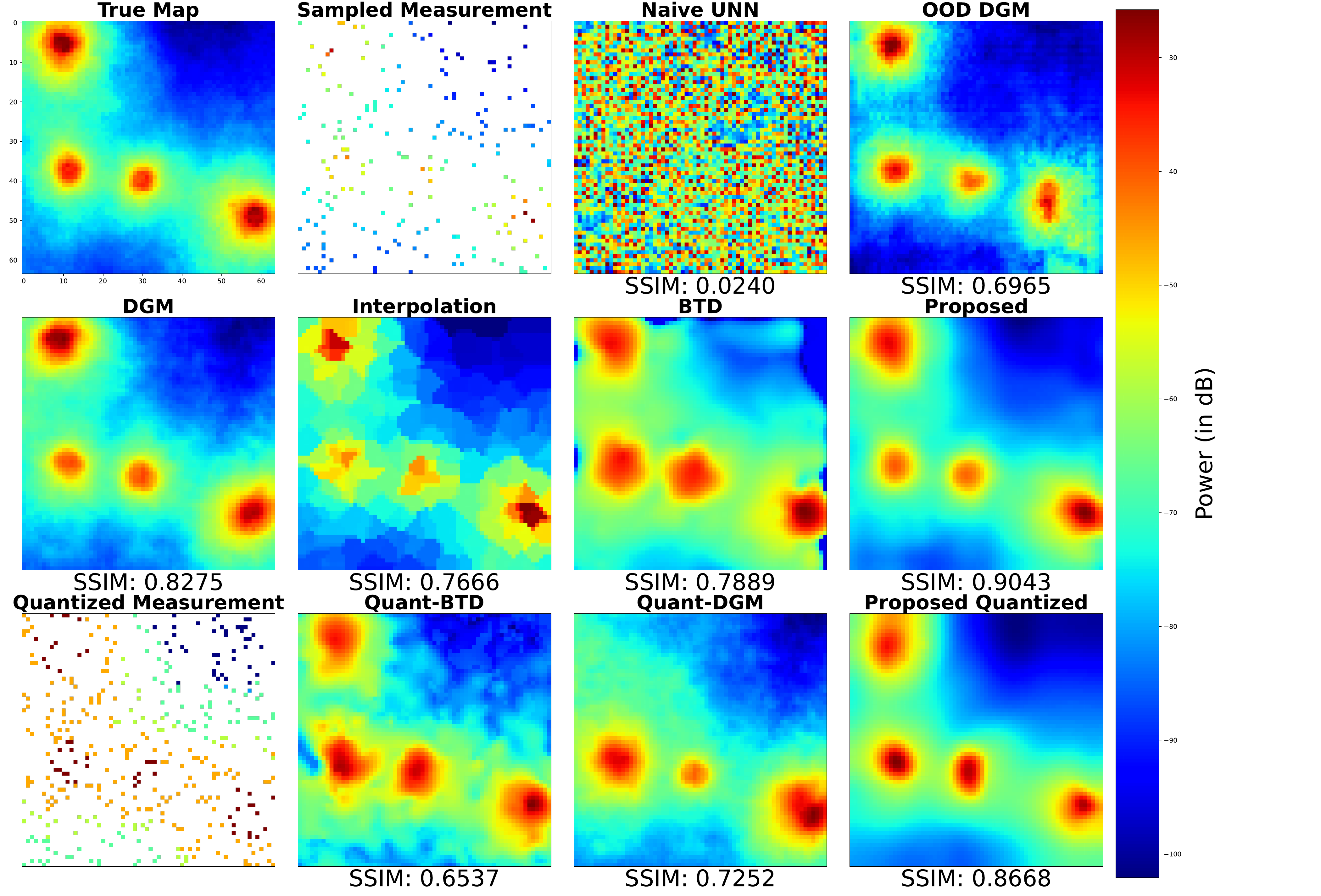}
    \caption{Rows 1-2: full-precision measurements ($\rho = 5 \%$). Row 3: quantized measurements using $B=3$ bits ($\rho = 10 \%$). $\eta= 6$, $X_c = 90$ for both cases.  }
    \label{fig:fp_viz} 
\end{figure}

\noindent
{\bf Data Generation}:
We consider a $64 \times 64$ spatial grid (1\,m$^2$ per cell) and $K = 64$ spectral bands.  
Radio maps are generated following \cite{zhang2020spectrum,shrestha2022deep,timilsina2023quantized}, using the standard SLF model with path loss and log-normal shadowing \cite{goldsmith2005wireless}.  
The shadowing is governed by the decorrelation distance $X_c$ and variance $\eta$---smaller $X_c$ and larger $\eta$ yield stronger shadowing. 
The radio maps of a wide range of scenarios can be simulated. For example, the urban area in the outdoor environment, typical values of the shadowing parameters are $X_c \in [20, 100]~$m and $\eta \in [4, 12]~$dB.
Each emitter’s PSD is synthesized by summing randomly scaled Gaussians \cite{zhang2020spectrum,shrestha2022deep,timilsina2023quantized}.

\noindent
{\bf Network Architecture}:
Our UNN has 3 layers with $3 \times 3$ convolution kernels and 6 channels per layer.  
The latent code $\bm z$ has dimension 16, and ${\cal G}_{\bm \theta}$ has 1080 trainable parameters.  
The detailed setup can be found in the supplementary material.

\noindent
{\bf Baselines}:
For full-precision cases, we compare with interpolation \cite{talvitie2015distance}, DGM \cite{shrestha2022deep}, BTD \cite{zhang2020spectrum}, and a naive UNN \cite{heckel2019deep} without spatio-spectral factorization. All UNN models use $1080 + 16R$ parameters for fairness. For quantized data, we compare with quantized BTD and DGM from \cite{timilsina2023quantized}.

\noindent
{\bf Performance Metric}:
We evaluate using \textit{structural similarity index measure} (SSIM) \cite{wang2004image} between the log-transformed ground-truth and recovered maps, averaged over all bands. SSIM ranges from 0 to 1, with higher being better.

\noindent
{\bf Results of Full-Precision Cases}:
Fig.~\ref{fig:fp_viz} (rows 1–2) shows recovery at the 30th band with $R=6$ emitters and $\rho = 5\%$. Our method achieves the best result. In contrast, the naive UNN \cite{heckel2019deep} underperforms, likely due to limited capacity. DGM performs well when trained on broad SLF ranges ($\eta\!\in\![4,8]$, $X_c\!\in\![40,90]$), but degrades under out-of-distribution (OOD) settings ($\eta\!\in\![7,8]$, $X_c\!\in\![40,70]$). Our method avoids this training-testing mismatch.
Table~\ref{table:merged_R} (cols. 2–6) shows results for varying $R$. Our method consistently outperforms others.

\begin{table}[t!]
    \begin{center}
    \caption{\texttt{SSIM} scores under various $R$'s. $\eta=6$, $X_c = 90$,  $\rho= 10\%$ for full-precision and $\rho=5\%$, $B=3$ bits for the quantized versions. \texttt{Q-} represents ``quantized version''.}
    \label{table:merged_R}
    \resizebox{\linewidth}{!}{ \Huge
    \begin{tabular}{|c | c | c | c | c | c || c | c | c | c}
        \toprule
    $R$ & \texttt{Naive UNN} & \texttt{Interp.} & \texttt{BTD} & \texttt{DGM} & \texttt{Prop.} & \texttt{Q-BTD} & \texttt{Q-DGM} & \texttt{Q-Prop.} \\ \midrule 
    1 & 0.6763 & 0.6943 & 0.7309 & 0.7509 & {\bf 0.8044} & 0.6745 & 0.7076 & {\bf 0.7987} \\
    2 & 0.1352 & 0.7355 & 0.7343 & 0.7520 & {\bf 0.8647} & 0.6421 & 0.7267 & {\bf 0.8375} \\
    3 & 0.0625 & 0.6825 & 0.7467 & 0.7544 & {\bf 0.8498} & 0.6250 & 0.7355 & {\bf 0.8061} \\
    4 & 0.0076 & 0.7625 & 0.7534 & 0.7591 & {\bf 0.8632} & 0.6338 & 0.7021 & {\bf 0.8571} \\
    5 & 0.0056 & 0.7595 & 0.8064 & 0.7675 & {\bf 0.8385} & 0.6039 & 0.7371 & {\bf 0.7845} \\
    6 & 0.0071 & 0.7370 & 0.7679 & 0.7315 & {\bf 0.8483} & 0.5413 & 0.7701 & {\bf 0.7906} \\
    \bottomrule
    \end{tabular}
    }
\end{center}
\end{table}

\noindent
\textbf{Results of Quantized Cases}:
Fig.~\ref{fig:fp_viz} (row 3) shows performance under 3-bit quantization, where our UNN remains competitive.  
Table~\ref{table:merged_R} (cols. 7–9) reports results for varying $R$, consistent with the full-precision case.

\noindent
{\bf Ablation Study}:  
Table~\ref{tab:ssim_vs_parameters} shows that modeling a $64\times64\times64$ radio map with a naive UNN requires about 30× more parameters than the proposed for similar accuracy.

Table~\ref{tab:combined_ablations} presents an ablation study on the proposed UNN components, e.g., layers, channels, kernel size, and decoder components. Using 3 layers, 6 channels, and a $3\times3$ kernel as baseline, we vary one parameter at a time. Results show this compact configuration achieves strong SSIM with fewer parameters. The Decoder Ablation column confirms that removing domain factorization, bilinear upsampling, or channel normalization degrades performance notably.

\noindent
{\bf Environment/Parameter Sensitivity}: 
Table~\ref{tab:shadow_ablation} shows performance under varying shadowing parameters, namely, decorrelation distance \(X_c\) and variance \(\eta\). As expected, lower \(X_c\) or higher \(\eta\) leads to degradation. However, the proposed method exhibits higher robustness relative to the baselines.

Table~\ref{tab:hyper_params} presents a sensitivity study on settings and key hyperparameters of our method, including initialization methods; defaults are in the supplementary material. Results show that the method is not sensitive to such parameters overall.

Table~\ref{tab:mismatch_r_combined} reports SSIM as the estimated number of emitters \(\widehat{R}\) varies when the ground-truth $R=4$. While classical model-order selection can estimate \(R\)~\cite{stoica2005spectral}, inaccurate estimates and mismatches are possible. Our method is robust to overestimation of \(R\), with only slight degradation. This is consistent with observations from prior approaches using \eqref{eq:radio_model} \cite{shrestha2022deep,zhang2020spectrum,timilsina2023quantized}.

Table~\ref{tab:mismatch_r_combined} also presents the runtime performance of the methods.
Notably, all methods require inference-time optimization, and the proposed UNN approach offers competitive accuracy with comparable runtimes. We should mention that the UNN method is initialized by BTD. Therefore, the result implies that it does not add significant runtime to its initialization stage yet substantially improves accuracy.

\begin{table}[t!]
\centering
\caption{ The numbers of parameters needed to model a $64\times 64\times 64$ radio map ($R=3, X_c=90, \eta=6$) by the proposed and naive UNNs, respectively. }
\label{tab:ssim_vs_parameters}
\resizebox{\linewidth}{!}{ \Huge
\begin{tabular}{|c|c c c c c c|c|}
\hline
\textbf{Model} & \multicolumn{6}{c|}{\textbf{Naive UNN (Params / SSIM)}} & \textbf{Proposed} \\ \hline
\textbf{\#Params} & 1080 & 3510 & 4810 & 6426 & 10206 & 30240 & 1080 \\ \hline
\textbf{SSIM}     & 0.7254 & 0.8215 & 0.8938 & 0.9392 & 0.9566 & 0.9656 & 0.9641 \\ \hline
\end{tabular} }
\end{table}

\begin{table}[t!]
\begin{center}
\caption{ Ablation study on architecture. $X_c=90$, $\eta=4$, $R=4$, $\rho=10\%$. $| \bm \theta |$ represents the cardinality of $\bm \theta$.}
\label{tab:combined_ablations}
\resizebox{\linewidth}{!}{\Huge
\begin{tabular}{|c|c||c|c||c|c||c|c|}
\hline
\multicolumn{2}{|c||}{\textbf{Layers}} & \multicolumn{2}{c||}{\textbf{Channels}} & \multicolumn{2}{c||}{\textbf{Kernel}} & \multicolumn{2}{c|}{\textbf{Decoder Ablations}} \\
\hline
\# Layers & SSIM & \# channels (\( |  \bm \theta | \)) & SSIM & Size & SSIM & Operation & SSIM \\
\hline
1 & 0.8478 & 3 (504) & 0.8567 & 2×2 & 0.3588 & w/o factorization & 0.1534 \\
3 & 0.8958 & 6 (1080) & 0.8958 & 3×3 & 0.8958 & w/o upsampling   & 0.7738 \\
5 & 0.8992 & 16 (7200)& 0.9003 & 4×4 & 0.8590 & w/o channel norm  & 0.8410 \\
7 & 0.8124 & 128 (444672)& 0.9055 & 5×5 & 0.8558 & proposed         & \textbf{0.8958} \\
\hline
\end{tabular} }
\end{center}
\end{table}

\begin{figure}[!t]
    \centering
    \includegraphics[width=\linewidth]{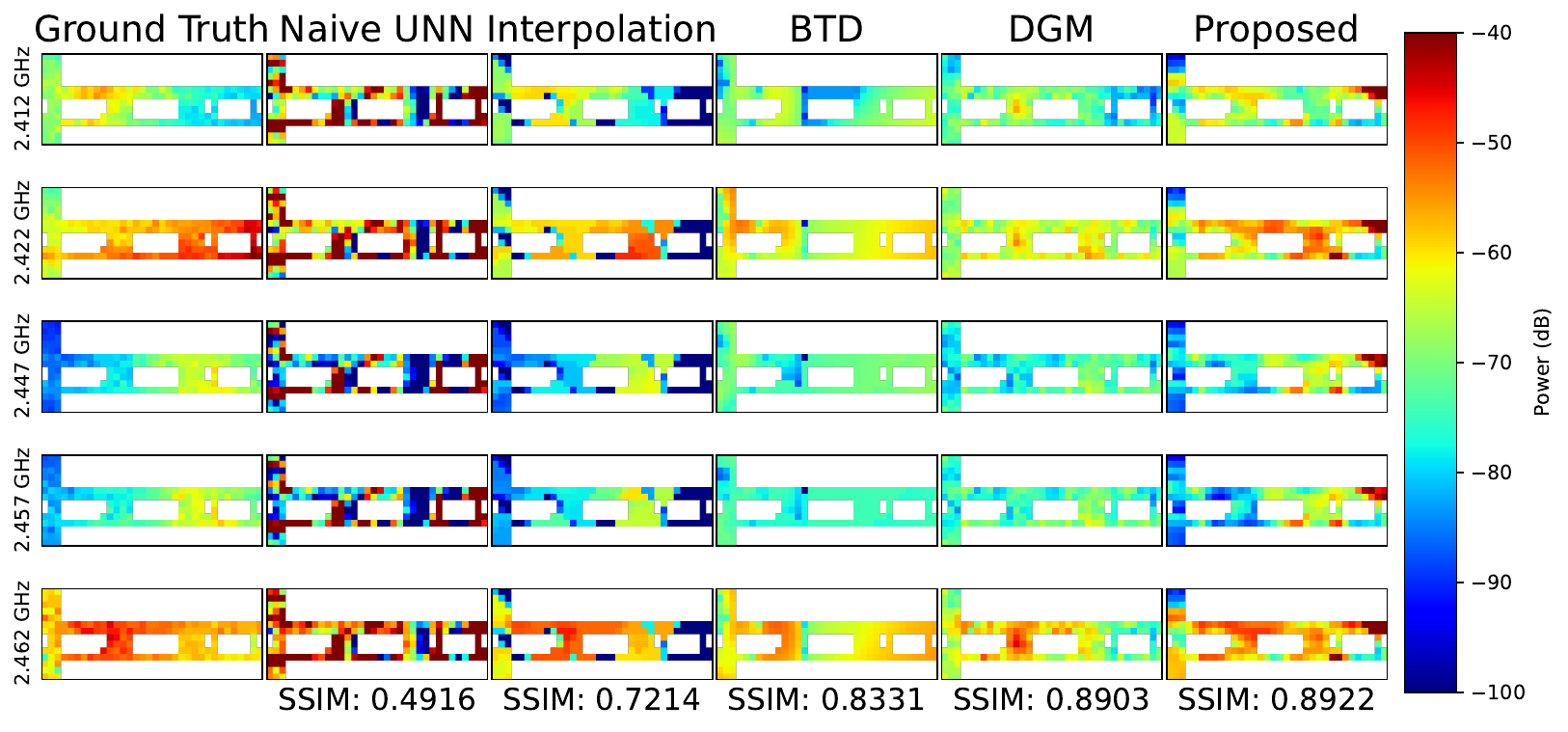}
    \caption{Real data experiment with $\rho = 10\%$.}
    \label{fig:real_viz}
\end{figure}

\noindent
\textbf{Real-Data Experiment}:
We use data from a $14 \times 34\,{\rm m}^2$ indoor space at Mannheim University over 9 frequencies \cite{realdataset}; see \cite{shrestha2022deep, timilsina2023quantized} for details. The setup is similar, with 4 intermediate channels and $R = 7$ \cite{zhang2020spectrum,shrestha2022deep,timilsina2023quantized}.  
Fig.~\ref{fig:real_viz} shows reconstructions across 5 frequencies, where our method outperforms baselines on this complex real-world data.

 Code and extra results are in \url{https://github.com/XiaoFuLab/Radio-Map-Estimation-Untrained-Neural-Network}.

\begin{table}[t!]
\centering
\caption{ SSIM scores under varying $X_c$ (left) and $\eta$ (right). $R=4$, $\rho=10\%$, default $\eta=4$ (left) and $X_c=90$ (right). }
\label{tab:shadow_ablation}
\resizebox{\linewidth}{!}{
\Huge
\begin{tabular}{|c|ccccc||c|ccccc|}
\hline
\multicolumn{6}{|c||}{Varying shadowing parameter $X_c$} & \multicolumn{6}{c|}{Varying shadowing parameter $\eta$} \\
\hline
$X_c$ & \texttt{Naive} & \texttt{Interp.} & \texttt{BTD} & \texttt{DGM} & \texttt{Prop.} & $\eta$ & \texttt{Naive} & \texttt{Interp.} & \texttt{BTD} & \texttt{DGM} & \texttt{Prop.} \\
\hline
30 & 0.2069 & 0.6593 & 0.6961 & \bf 0.7512 & 0.7470 & 4 & 0.3913 & 0.7826 & 0.7625 & 0.7813 & \bf 0.8835 \\
50 & 0.3761 & 0.7417 & 0.7582 & 0.7412 & \bf 0.8017 & 5 & 0.3481 & 0.7273 & 0.7099 & 0.7060 & \bf 0.8341 \\
70 & 0.1532 & 0.7413 & 0.7807 & 0.7561 & \bf 0.8109 & 6 & 0.3502 & 0.7036 & 0.6824 & 0.6866 & \bf 0.7910 \\
90 & 0.3133 & 0.7641 & 0.7423 & 0.7620 & \bf 0.8626 & 7 & 0.2212 & 0.7117 & 0.7714 & 0.7358 & \bf 0.7768 \\
100 & 0.1314 & 0.7950 & 0.8458 & 0.8416 & \bf 0.8701 & 8 & 0.4035 & 0.7410 & 0.7412 & 0.6755 & \bf 0.7764 \\
\hline
\end{tabular} }
\end{table}

\begin{table}[t!]
    \begin{center}
    \caption{ \texttt{SSIM} under various hyperparameters. \(X_c=90\), \(\eta=4\), $R=4$ and \(\rho=10\%\).}
    \label{tab:hyper_params}
    \resizebox{1.0\linewidth}{!}{ \Huge 
    \begin{tabular}{|c | c || c | c || c | c || c | c || c | c || c | c|}
        \toprule
    $\lambda_1$ & SSIM  & $\lambda_2$ & SSIM  & $\alpha$ & SSIM  & $\beta$ & SSIM  & Initialization & SSIM  \\ \midrule 
    0.0001 & 0.8924 & 0.0001 & 0.8896 & 0.005 & 0.8101 & 0.0001 & 0.8802 & {\rm unif}[-1, 1] & 0.8688 \\
    0.001  & 0.8947 & 0.001  & 0.8947 & 0.01 & 0.8830 & 0.001 & 0.8947 & Xavier \cite{glorot2010understanding} & 0.8652 \\
    0.01   & 0.8907 & 0.01   & 0.8876 & 0.05 & 0.8947 & 0.01 & 0.8766 & BTD & 0.8947 \\
    0.1    & 0.8864 & 0.1    & 0.8885 & 0.5 & 0.8735 & 0.1 & 0.8561 & Interpolation & 0.8892 \\
    \bottomrule
    \end{tabular} }
\end{center}
\end{table}

\begin{table}[t!]
    \begin{center}
    \caption{ \texttt{SSIM} scores with (runtime in seconds) under various $\widehat{R}$. Ground-truth $R=4$, $X_c=90$, $\eta=4$, $\rho=10\%$ for full-precision measurement. }
    \label{tab:mismatch_r_combined}
    \resizebox{0.95\linewidth}{!}{ 
    \begin{tabular}{|c|c|c|c|}
        \toprule
        $\widehat{R}$ & \texttt{BTD} & \texttt{DGM} & \texttt{Proposed} \\
        \midrule
        1 & 0.6434 (1.9679 s) & 0.5693 (2.4621 s) & \textbf{0.7393} (2.3422 s) \\
        2 & 0.7591 (2.0334 s) & 0.7268 (2.5104 s) & \textbf{0.8530} (2.3646 s) \\
        3 & 0.7770 (2.1399 s) & 0.7713 (2.5768 s) & \textbf{0.8716} (2.4180 s) \\
        4 & 0.8086 (2.2294 s) & 0.8120 (2.5449 s) & \textbf{0.8983} (2.4168 s) \\
        5 & 0.7809 (2.3598 s) & 0.8109 (2.5509 s) & \textbf{0.9053} (2.4455 s) \\
        6 & 0.7671 (2.4732 s) & 0.8098 (2.5650 s) & \textbf{0.8930} (2.4408 s) \\
        7 & 0.7484 (2.6985 s) & 0.7574 (2.5220 s) & \textbf{0.8830} (2.4895 s) \\
        \bottomrule
    \end{tabular} }
\end{center}
\end{table}

\section{Conclusion and Future Direction}

This letter addressed the SC problem using a custom-designed untrained neural network (UNN) as a deep prior.  
By leveraging physical properties of radio maps, our architecture reduces parameter count while maintaining expressiveness, leading to improved sample efficiency.  
The proposed method performs competitively with state-of-the-art approaches, such as learned DGM-based models, but avoids issues like training-testing mismatch.
 As a future direction, incorporating advanced propagation effects (e.g., multipath, diffraction, scattering) into the UNN architecture or loss in a differentiable way could enhance modeling accuracy and interpretability in complex settings.

\clearpage

\bibliographystyle{ieeetr}
\bibliography{refs}

\clearpage

\appendices
\input{appendix}

\end{document}

%% file: def.tex
\usepackage{steinmetz}

\usepackage{amsmath}
\usepackage{graphicx}

\newcommand{\bc}{\boldsymbol{c}}

\newcommand{\g}{\boldsymbol{g}}

\newcommand{\z}{\boldsymbol{z}}

\newcommand{\C}{\boldsymbol{C}}

\renewcommand{\S}{\boldsymbol{S}}

\newcommand{\U}{\boldsymbol{U}}

\newcommand{\Y}{\boldsymbol{Y}}
\newcommand{\Z}{\boldsymbol{Z}}

\newcommand{\btheta}{ {\boldsymbol{\theta}} }

\newcommand{\bTheta}{ { \boldsymbol{\Theta} } }

\newcommand{\tM}{\underline{ \boldsymbol{M} }}
\newcommand{\tN}{\underline{ \boldsymbol{N} }}

\newcommand{\tX}{\underline{ \boldsymbol{X} }}
\newcommand{\tY}{\underline{ \boldsymbol{Y} }}

\newcommand{\gtX}{\underline{\boldsymbol{X} }^\natural}

\newcommand{\cC}{\mathcal{C}}

\newcommand{\cH}{\mathcal{H}}

\newcommand{\cL}{\mathcal{L}}
\newcommand{\cM}{\mathcal{M}}
\newcommand{\cN}{\mathcal{N}}

\newcommand{\cQ}{\mathcal{Q}}

\newcommand{\cX}{\mathcal{X}}

\newcommand{\cZ}{\mathcal{Z}}

\newcommand{\bbR}{ {\mathbb{R}} }

\usepackage{textcomp}

\makeatletter
\newcommand{\vast}{\bBigg@{4}}
\newcommand{\Vast}{\bBigg@{5}}
\makeatother

\DeclareMathOperator*{\minimize}{\textrm{minimize}}

\usepackage{xcolor}
\usepackage{psfrag,framed}
\usepackage{lipsum}
\PassOptionsToPackage{normalem}{ulem}
\usepackage{ulem}

\definecolor{shadecolor}{RGB}{220,220,220}
\usepackage{caption}
\usepackage{subcaption}

%% file: appendix.tex
\begin{center}
    {\normalsize {\bf Supplementary Material of ``Domain-Factored Untrained Deep Prior for Spectrum Cartography''}}
    
    {Subash Timilsina, Sagar Shrestha, Lei Cheng, and Xiao Fu}
\end{center}

\section{Proof of Lemma~\ref{lemma:x_unn}}\label{app:cover}
To prove Lemma~\ref{lemma:x_unn}, we first consider the following lemma:
\begin{Lemma}\label{lemma:network} Assume that ${\cal G}_{\btheta}(\z)$ follows the structure in \eqref{eq:deep_dec_arch}, and that each activation is $\rho_i$-Lipschitz continuous $\forall i \in [L]$. Also, define 
    $\cH =  \{ {\cal G}_{\btheta}(\z) ~|~ {\btheta} \in {\cal M}, \z \in \cZ \}$
 where 
$\cM = \{ \btheta = [\bm \Theta^0 \ldots \bm \Theta^L] ~|~ \bm \Theta^{i} \in \bbR^{k_i \times k_{i+1}}, \| \bm \Theta_{i} \|_{\sigma} \leq s$, 
$\| \bm \Theta_i \|_{2, 1} \leq b  \}$ and $\cZ = \{ \z \in \bbR^{D_0^2} ~|~ \| \z \|_{2} \leq a \}.$
Let $W := \max_{i=1}^{L}( k_i )$ denote the maximum layer width. Then we have
    \begin{align}
   \log ( {\sf N}(\cH, \epsilon) ) \leq & \nicefrac{4 a^2 b^2 P \log(2 W^2) s^{2L - 2} L^3}{\epsilon^2} 
   + \text{\tiny $D_0^2 \log$} \left(  \nicefrac{6 P a}{\epsilon} \right ) \label{eq:cover_deep_decoder}
\end{align}
\end{Lemma}

\begin{IEEEproof}
Let us denote $\cH_{\z} = \{{\cal G}_{\btheta}(\z) ~|~ \btheta \in {\cal M}\}$. 
For any $\epsilon > 0$, let $\overline{\cH}_{\z}$ be the $\frac{\epsilon}{2}$-net of $\cH_{\z}$. Then, according to \cite[Theorem.~3.3]{bartlett2017spectrally}, the following holds:
\begin{align} \label{eq:cover_network}
     \log ( |\overline{\cH}_{\z} |  ) \leq \nicefrac{4 a^2 b^2 P \log(2 W^2) s^{2L - 2} L^3}{\epsilon^2},
\end{align}
where $P = \prod_{i = 1}^{L}\rho_{i}^{2}.$ Similarly, let $\overline{\cZ}$ be the $\frac{\epsilon}{2P}$-net of $\cZ$. Then, one can bound $|\overline{\cZ}|$ as follows \cite{shrestha2022deep}:
\begin{align} \label{eq:cover_latent}
    \log ( |\overline{\cZ} |  ) \leq D_0^2 \log \left(  \nicefrac{6 P a}{\epsilon} \right ).
\end{align}
For any ${\cal G}_{\btheta}(\z) \in \cH_{\z}$ and $\z \in {\cal Z}$, let ${\cal G}_{\btheta'}(\z) \in \overline{\cH}_{\z}$ and $\z' \in \overline{\cZ}$ be such that $\| {\cal G}_{\btheta}(\z) - {\cal G}_{\btheta'}(\z) \| \leq \nicefrac{\epsilon}{2}$ and $\| \z - \z'\|_2 \leq \nicefrac{\epsilon}{2P}$. Then, the following chain of inequalities hold:
\begin{align*}
    \text{\small $\| {\cal G}_{\btheta}(\z) - {\cal G}_{\btheta'}(\z') \|_{\rm F}$ }
    &\text{\small $\stackrel{(a)}{\leq} \| {\cal G}_{\btheta}(\z) - {\cal G}_{\btheta'}(\z) \|_{\rm F} + \| {\cal G}_{\btheta'}(\z) - {\cal G}_{\btheta'}(\z') \|_{\rm F}$ }\\
    & \leq \nicefrac{\epsilon}{2} + P \| \z - \z'\|_2  \leq \nicefrac{\epsilon}{2} + P \nicefrac{\epsilon }{2P} \leq \epsilon,
\end{align*}
where $(a)$ is due to triangle's inequality.  
Hence, the set $\overline{\cal H} = \{{\cal G}_{\btheta}(\z) \in \overline{\cal H}_{\z} ~|~ \z \in \overline{\cal Z}\}$ is an $\epsilon$-cover of $\cH$. Since $|\overline{\cal H}| = |\overline{\cal H}_{\z}| |\overline{\cal Z}|$, $\log ( {\sf N}(\cH, \epsilon) )$ can be upper-bounded by summing \eqref{eq:cover_network} and \eqref{eq:cover_latent}. This completes the proof.
\end{IEEEproof}

 The covering number of ${\cal X}^{\rm UNN}$ can be upper bounded using \eqref{eq:cover_deep_decoder} and the covering number of $\cC = \{ \C ~|~ \C \geq \bm 0, \| \bm c_r \|_2 \leq \kappa, \forall r \in [R] \}$. Then, using Lemma \ref{lemma:network} and following the same steps from proof of Lemma~2 \cite{shrestha2022deep, timilsina2023quantized}  leads to
 \begin{align*}
     \log ( {\sf N}(&{\cal X}^{\rm UNN}, \epsilon) ) \leq  \nicefrac{R^3 (\kappa + \gamma) a^2 b^2 P \log(2 W^2) s^{2L - 2} L^3}{\epsilon^2} \nonumber \\
     &+ R D_0^2 \log \left(  \nicefrac{6 R P a (\kappa + \gamma)}{\epsilon} \right ) + R K \log \left( \nicefrac{3 R \kappa (\kappa + \gamma)}{\epsilon} \right).
 \end{align*}

 \section{The Quantized Measurement Case}\label{app:quant}
In this setting, we assume the sensors use the quantizer ${\cal Q}(\cdot):\mathbb{R}\rightarrow \mathbb{Z}$ that maps the PSD at $(i,j,k)$ to an integer using Gaussian quantization strategy \cite{bhaskar2016probabilistic, cao2015categorical, ghadermarzy2018learning}:
\begin{align}    \label{eq:model_qsc}
    &\tY_q(i,j,k) = \cQ(h( \gtX(i,j,k) ) +\tN(i,j,k) )\\
    &\cQ(x) = \ell, \text{ if } b_{\ell-1} < x \leq b_{\ell}, \quad \ell \in [L] , \nonumber
\end{align}
where $\tN(i,j,k) \sim \cN(0,\sigma^2)$, and $\{b_\ell\}_{\ell=1}^L$ are the quantization bins; see \cite{timilsina2023quantized} for definition of $h$.
Then, we tackle the following MLE \cite{timilsina2023quantized}:
\begin{align}\label{eq:quantized_criterion}
 &\underset{\btheta, \Z, \C \geq \bm 0 }{\minimize} \sum_{(i,j) \in \bm \varOmega}\sum_{k=1}^K \sum_{q=1}^Q \mathbb{1}_{[{\tY_q(i,j,k) = q}]} \cL^{\rm quant}_{(i, j, k)}
\end{align}
where $\cL^{\rm quant}_{(i, j, k)} = \log ( f_{q}  \big(h ( [ \sum_{r=1}^R \g_{\btheta}(\z_{r}) \circ \bc_r  ]_{i,j,k} ) \big) )$.

\begin{Prop}\label{prop:quantized}
        Assume that $\bm \varOmega$ is uniformly sampled from $[I] \times [J]$ with replacement such that $|\bm \varOmega| = N$ and sensor follows quantization as described in \eqref{eq:model_qsc}. Suppose that $\tX^{\ast} = \sum_{r=1}^R {{\cal G}_{\btheta^\star}(\z_r^{\ast})} \circ {\bc}_r^{\ast}$, where $(\bm \theta^\star,\{\bm z_r^\star,\bm c_r^\star\}_{r=1}^R)$ is any optimal solution of \eqref{eq:quantized_criterion} and $\tX^\star\in {\cal X}^{\rm UNN}$.
      Then the following holds with probability of at least $1 - 2\delta$: 
            \begin{align*}
                \frac{\|\tX^{\star} - \tX_{\natural}\|_{\rm F}}{\sqrt{IJK} } \leq {\sf O} \left ( \frac{\sqrt{R}}{K\sqrt{N}} + \sqrt{\frac{ ( \log ( {\sf N}({\cal X}^{\rm UNN}, \epsilon) ) )}{N}} \right ) + \nu^{\rm UNN} ,
            \end{align*}
        where, $\nu^{\rm UNN} =\min_{{\tX}\in \cX^{\rm UNN}}\|\tX - \gtX \|_{\infty} $.
\end{Prop}
The proof is by substituting the result of Lemma~\ref{lemma:x_unn} into the proof of \cite[Theorem~2]{timilsina2023quantized}.

\section{Algorithm and Network Details}\label{app:algo}
The unified algorithm for both Euclidean and quantized measurements is summarized in Algorithm \ref{algo:fp}, where ${\cal L}$ represents the objective in \eqref{eq:full_precision} or \eqref{eq:quantized_criterion} 
with regularization  $\lambda_1 \|\bm Z\|_{\rm F}^2$, $\lambda_2 \|\bm C\|_{\rm F}^2$, and $\lambda_3 \|\bm \Theta^{i}\|_{\rm F}^2, \forall \ell \in [L]$. 
We set $\lambda_1 = 0.001$, $\lambda_2 = 0.001$, and $\lambda_3 = 0.0001$.
We stop the iterations if the relative change of loss is less than $\epsilon = 10^{-3}$ or $ r = {\rm MaxIter}: = 300$ is reached. We use the Adam optimizer \cite{kingma2014adam} for the updates in Line 6 and 7, where $\overline{\nabla}$ denotes a gradient-based direction for updating. We set the learning rates of UNN parameters and $\C$ to be $\alpha=0.05$ and $\beta=0.001$, respectively.

The architecture of ${\cal G}_{\btheta}$ is summarized in Table~\ref{tab:network_params}. Each \texttt{Conv2d}($k_i \rightarrow k_{i+1}$, $n_{k_i} \times n_{k_i}$) maps $k_i$ input channels to $k_{i+1}$ outputs using an $n_{k_i} \times n_{k_i}$ kernel.

\begin{algorithm}[!t] 
\footnotesize
\caption{Proposed Algorithm} 
\label{algo:fp}
\KwData{$\tY, \tM_{\bm \Omega}, R, \epsilon, \alpha, \beta,$ MaxIter and $\cL$ is \eqref{eq:full_precision} or \eqref{eq:quantized_criterion}.}
\KwResult{$\widehat{\tX} = \sum_{r=1}^R {\g_{\btheta}(\z_r)}^{(k)} \circ {\bc}_r^{(k)}$}
Initialize $\Z, \btheta$ such that $\sum_{r=1}^R {\g_{\btheta}(\z_r)} \circ {\bc}_r \approx \widehat{\tX}^{\rm BTD}$ and $\C \sim \U[0,1]$;

$k\leftarrow 1$;

\While{$|\cL(\btheta^{(k)}, \Z^{(k)}, \C^{(k)}) - \cL(\btheta^{(k-1)}, \Z^{(k-1)}, \C^{(k-1)}) |\geq \epsilon$ or $k \leq $ MaxIter}{
        
     $\C^{(k+1)} \gets \C^{(k)} - \alpha \overline{\nabla}_{\C}  \cL(\btheta^{(k)}, \Z^{(k)}, \C^{(k)})$;

     $\C^{(k+1)} \gets {\rm max}( \C^{(k+1)}, \mathbf{0} )$; 

     $\Z^{(k+1)} \gets \Z^{(k)} - \beta \overline{\nabla}_{\Z} ~ \cL(\btheta^{(k)}, \Z^{(k)}, \C^{(k)})$;

     $\btheta^{(k+1)} \gets \btheta^{(k)} - \beta \overline{\nabla}_{\btheta} ~ \cL(\btheta^{(k)}, \Z^{(k)}, \C^{(k)})$;
    
     $k \leftarrow k + 1$;
}
\end{algorithm}

\begin{table}[t!]
\centering
\caption{ UNN architecture. \texttt{UpBlock} = \texttt{Upsample} + \texttt{ReLU} + \texttt{BatchNorm}. }
\label{tab:network_params}
\resizebox{\linewidth}{!}{
 \Huge
\begin{tabular}{|l|l|c|c|}
\hline
\textbf{Module} & \textbf{Operations} & \textbf{\#Params} & \begin{tabular}[c]{@{}c@{}}\textbf{Output} \\ $(k_i \times D_i \times D_i)$\end{tabular} \\ \hline
Input & -- & 0 & $(1, 4, 4)$ \\ \hline
Input layer & \texttt{Conv2d(1$\rightarrow$6,$3\times3$)} + \texttt{UpBlock} & 66 & $(6, 8, 8)$ \\ \hline
Hidden 1 & \texttt{Conv2d(6$\rightarrow$6,$3\times3$)} + \texttt{UpBlock} & 336 & $(6, 16, 16)$ \\ \hline
Hidden 2 & \texttt{Conv2d(6$\rightarrow$6,$3\times3$)} + \texttt{UpBlock} & 336 & $(6, 32, 32)$ \\ \hline
Hidden 3 & \texttt{Conv2d(6$\rightarrow$6,$3\times3$)} + \texttt{UpBlock} & 336 & $(6, 64, 64)$ \\ \hline
Output & \texttt{Conv2d(6$\rightarrow$1,$1\times1$)} + \texttt{Sigmoid} & 6 & $(1, 64, 64)$ \\ \hline
\multicolumn{2}{|l|}{\textbf{Total trainable parameters}} & \textbf{1080} & \\ \hline
\end{tabular}} 

\end{table}